\documentclass[]{article}
\usepackage{amssymb}
\usepackage{amsmath}
\newcommand{\benumerate}{\begin{enumerate}}
\newcommand{\eenumerate}{\end{enumerate}}

\newcommand{\bitemize}{\begin{itemize}}

\newcommand{\eitemize}{\end{itemize}}

\input{tcilatex}

\begin{document}

\title{Unified treatment of heterodyne detection: \\ 
the Shapiro-Wagner and Caves frameworks}

\author{Giulio Landolfi \thanks{
e-mail: Giulio.Landolfi@le.infn.it} \ ,
Giovanna Ruggeri\thanks{
 e-mail: Giovanna.Ruggeri@le.infn.it} \
and Giulio Soliani\thanks{
e-mail: Giulio.Soliani@le.infn.it} \\
{\em Dipartimento di Fisica dell'Universit\`{a} di Lecce}\\
{\em and INFN, Sezione di Lecce, I-73100 Lecce, Italy} }
\date{\today}
\maketitle

\begin{abstract}

A comparative study is performed on two heterodyne systems of photon detectors expressed in terms of a 
signal annihilation operator and an image band creation operator called Shapiro-Wagner and Caves' frame, 
respectively. This approach is based on the introduction of a convenient operator $\hat \psi$ which allows 
a unified formulation of both cases. For the Shapiro-Wagner scheme, where $[\hat \psi, \hat \psi^{\dag}] =0$, 
quantum phase and amplitude are exactly defined in the context of relative number state (RNS) representation, 
while a procedure is devised to handle suitably and in a consistent way Caves' framework, characterized by 
$[\hat \psi, \hat \psi^{\dag}] \neq 0$, within the approximate simultaneous measurements of noncommuting 
variables. In such a case RNS phase and amplitude make sense only approximately.



\end{abstract}

\section{Introduction}

The problem of defining appropriately the phase of an electromagnetic field at the quantum level has a long story, 
which was first addressed by Dirac in 1927 \cite{D}. 

In this work we do not touch this historical excursus, although it should be keenly interesting for different 
reasons, but we limit ourselves to outline the main aspects on the topic exhaustively handled in the basic 
reviews of Carruthers-Nieto in 1968 \cite{CN} and Lynch in 1995 \cite{L}.

A consistent progress versus a better statement of the phase problem is constituted by the relative 
number state (RNS) representation devised by Ban \cite{B2} some years ago. 
The RNS phase operator formulation is strictly connected with the Liouville space formulation and 
to thermofield dynamics (see Ref. \cite{B2} and references therein). This formulation is particularly 
useful for studying the number-phase quantization in the Josephson junctions with ultrasmall capacitance. 
For the investigation of this type of quantization in the mesoscopic Josephson junctions the 
RNS representation reveals adequately profitable \cite{B2}. For completeness, we shall briefly 
summarize the principal steps of the Ban procedure. Pertinent papers are also Refs. \cite{HRADIL} 
and \cite{B4}. Consider a system composed of two independent and distinguishable subsystems $A$ and $B$. 
Then, let ${\cal H = \cal H_{A} \otimes \cal H_{B}}$ be the extended Hilbert space and $\hat a$, 
$\hat b$ the annihilation operators involved in the Shapiro-Wagner (SW) operator \cite{SW}

\begin{equation}
\label{GG1}
 {\hat Y_{SW}}= \hat a + \hat b^{\dag}.
\end{equation}

The subsystems $A$ and $B$ can be described by the complete orthogonal discrete bases 
of Fock states $|m>_{A}\otimes |n>_{B}$, $m,n=0,1, ...$, or alternatively by the RNS 
basis generated by the states \cite{HRADIL}

\begin{equation}
\label{GG2}
|n,m>>=\Theta (n) |m+n>_{A}|m>_{B}+\Theta (-n -1)|m>_{A}|m-n>_{B},
\end{equation}
where $-\infty <n< \infty $, $m\ge 0 $, the function $\Theta (n)$ is  given by
$\Theta (n) = 1$ for $n\ge 0$ and $\Theta (n) =0$ for $n<0$.

The states (\ref{GG2}) correspond to the product of Fock states defined on the Hilbert subspaces 
$\cal H_{A}$ and $\cal H_{B}$ on the ground of the condition

\begin{equation}
\label{GG3}
|n-m, min(m,n)>>=|m>_{A}|n>_{B}.
\end{equation}

Moreover, the number-difference operator $\hat N=\hat a^{\dag} \hat a - \hat b^{\dag} \hat b $ 
acting on the RNS $|n, m>>$ provides

\begin{equation}
\label{GG4}
\hat N |n,m>> = n|n,m>>.
\end{equation}

It can be shown that (see Refs. \cite{B2}, \cite{HRADIL}, \cite{B4}) 
the basis of the RNS's is complete and orthonormal. As a consequence, an unitary phase operator 
$\hat D \equiv \hat D_{RNS}$  exists on the Hilbert space $\cal H= \cal H_{A}\otimes \cal H_{B}$ 
in such a way that
\begin{equation}
\label{GG5}
\hat D = \sum_{m=0}^{\infty} \sum_{n=-\infty}^{\infty} |n-1,m>><<m,n|,
\end{equation}
obeying the relation
\begin{equation}
\label{GG6}
\hat D \hat D^{\dag}= \hat D^{\dag} \hat D = \hat 1
\end{equation}
and the commutation rule
\begin{equation}
\label{GG7}
[\hat D, \hat N ]= \hat D.
\end{equation}

Ban \cite{B2} points out that the property of the operator $\hat N $ allows to define the operator 
$\hat D $. This is due to the fact that the spectrum of $\hat N $ is unbounded. Thus, the introduction 
of the RNS representation makes it possible to define the operator $\hat D$, which plays the role of 
a phase operator as it is clarified in Ref. \cite{B2}.
According to Ban \cite{B4}, two ways can be adopted to define quantum mechanical phase operators. 
One is based on the polar decomposition of the annihilation operator of a photon (ideal phase), 
and the other on the use of phase-measurement processes (feasible phase). 
The relationship between the ideal and feasible phases is discussed in Refs. \cite{B4} and \cite{HRADIL}.
As it is noticed by Ban \cite{B4}, the RNS representation fits fairly with the (feasible) phase concept 
of Shapiro-Wagner \cite{SW}. Hradil \cite{HRADIL} has shown that the SW feasible phase is well described 
by a unitary phase operator (the SW phase operator)

\begin{equation}
\label{GG8}
\hat D_{SW}=\hat Y_{SW} (\hat Y_{SW}^{\dag} \hat Y_{SW})^{\frac{-1}{2}}.
\end{equation}
$\hat D_{SW}$ and $ \hat N $ satisfy the same commutation relation as for $\hat D \equiv \hat D_{RSN}$ 
(see Ref. \cite{HRADIL}). The RNS phase operator $\hat D \equiv \hat D_{RNS}$ is related to the SW 
phase operator \cite{HRADIL} by

\begin{equation}
\label{GG9}
\hat D_{RNS}=\hat U \hat D_{SW} \hat U^{\dag}
\end{equation}
where $\hat U$ is a nonunitary operator given by

\begin{equation}
\label{GG10}
\hat U = \sum_{n=-\infty}^{\infty} \int^{\infty}_{0} d\mu |n,[\mu]>><\mu,n|
\end{equation}
with $\hat U \hat U^{\dag} =\hat 1 $ and $\hat U^{\dag} \hat U \neq \hat 1$.

Here $|n, [\mu]>>$ is the relative-number state, $[\mu]$ is the integer part of $\mu$, and $|n, \mu>$ 
is the Lindblad-Nagel state \cite{LINDNAGEL} which is an element of the continuous basis for irreducible 
unitary representation of the $su(1,1)$ Lie algebra.
In the SW approach to the phase and squared-amplitude measurements, the squared-amplitude and phase 
information obtained from heterodyne detection are expressed by
\begin{equation}
\label{GG11}
V=|\hat Y_{SW}|^{2}
\end{equation}
and
\begin{equation}
\label{GG12}
\hat \Phi = arg (\hat Y_{SW}),
\end{equation}
where in (\ref{GG1}) $\hat a \equiv \hat a_{S}$ is identified by an annihilation operator of a photon 
{\it signal} mode and $ \hat b^{\dag} \equiv \hat a_{I}^{\dag}$ by a creation operator of an {\it image} 
mode (see the schematic description of Fig. 1 of the heterodyne apparatus employed by Shapiro-Wagner \cite{SW}).
For the discussion of many conceptual and technical aspects, such as for example the statistics of $V$ 
and $\hat \Phi $, and the role of uncertainties in optical heterodyne detection, we remind the reader 
to consult the original SW article \cite{SW}.
Anyway, we remark that in the SW strategy, what is essential is the commutation rule

\begin{equation}
\label{GG13}
[\hat Y_{SW}, \hat Y_{SW}^{\dag}]=0,
\end{equation}
as one can promptly establish starting from (\ref{GG1}).

However, a quite different physical context can be accomplished in the case in which the heterodyne detector 
is not of the SW type, but rather a power-detector corresponding to a measurement operator taking the form 
(Caves' operator)
\begin{equation}
\label{GG14}
\hat Y_{C}=(1+\frac{\nu_{IF}}{\nu_{0}})^{\frac{1}{2}}{\hat a}+(1-\frac{\nu_{IF}}{\nu_{0}})^{\frac{1}{2}}{\hat b^{\dag}}
\end{equation}
$(\hat a \equiv \hat a_{S}, \hat b^{\dag } \equiv \hat a_{I}^{\dag})$, which does not commute with its 
adjoint $\hat Y_{C}^{\dag}$. In other words, in the case of Shapiro-Wagner, phase and squared-amplitude can 
be simultaneously measured (see Eq. (\ref{GG13})), while for
\begin{equation}
\label{GG15}
[\hat Y_{C}, \hat Y_{C}^{\dag}]\neq 0
\end{equation}
simultaneous amplitude and phase measurements are not possible.
The configurations characterized by Eq. (\ref{GG14}) with property (\ref{GG15}) have been investigated by Caves 
\cite{CAVES} (see also Ref. \cite{YUEN}) and the inherent mechanism is called heterodyne with square-law detector 
(see Ref. \cite{SW} and references therein). In formula (\ref{GG14}), $\nu_{IF}$ is named intermediate frequency 
and $\nu_{0}$ is an optical frequency. The terminology used in some description of heterodyne detectors can 
be found, for instance, in Ref. \cite{PRATT}.

This paper has essentially a speculative character, as we shall illustrate below, addressed to getting a theoretical 
insight into two fundamental heterodyne frames of photon detectors with different characteristics. Precisely, 
the apparatus analyzed by Shapiro-Wagner \cite{SW} and that relative to Caves' scheme \cite{CAVES}.
These devices realize two opposite physical situations, which are handled here by resorting to a compact and 
unifying approach outlined in Section 3.
There two operators, $\hat \psi (t)$ and $\hat \psi^{\dag} (t)$ (see Eqs. (\ref{II1}) and (\ref{II2})), have been 
introduced depending on a linear combination of an annihilation signal field operator $\hat a(t)$ and an image 
band mode creation operator $\hat b^{\dag}$, via two  arbitrary constant real parameters $A$ and $B$.
The realization of quantum measurements (of phase and squared-amplitude) pertinent to the SW frame is based on 
the operator $\hat \psi_{SW}=\sqrt{A}\tilde a + \sqrt{B} \tilde b^{\dag}$ with $A=B$ (see Eq. (\ref{II12})) 
where $\hat \psi_{SW}$, $\hat \psi_{SW}^{\dag}$ commute. On the contrary, the theory of Caves' situation 
corresponds to the operator $\hat \psi_{C}=\sqrt{A}\tilde a + \sqrt{B} \tilde b^{\dag}$ with $A \neq B$ 
where $\hat \psi_{C}$, $\hat \psi_{C}^{\dag}$ do not commute. [The symbols $\tilde a $ and $\tilde b^{\dag}$ 
are defined in Sec. 3 and stand for rotation-valued operator representations of $\hat a$ and $\hat b^{\dag}$].
The comparative study of Shapiro-Wagner and Caves' frameworks constitutes the main purpose of this paper.

The foremost results achieved in our approach are:

i) a quantum extension is proposed of a phase concept following the track previously developed in 
\cite{NOVEL} for a generalized classical oscillator (see Sec. 2). This phase turns out to be 
self-adjoint and is particularly suitable in the context of SW heterodyne framework. Within the 
same context, a squared-amplitude can be defined in terms of quadrature components (see Eq. (\ref{Z16}) 
and Sec. 7). An interesting property of quadrature operators $\hat y_{1}$, $\hat y_{2}$ is given by Eqs. 
(\ref{Z17}) and (\ref{Z18}), where $\hat y_{1}$, $\hat y_{2}$ are expressed via $\hat \cos\theta$, 
$\hat \sin\theta$, and the generator of the parabolic SU(1,1) subgroup.

ii) Taking account of the content of Section 2, in Section 3 a unified formulation is set up of the 
Shapiro-Wagner and Caves' frames by means of the introduction of the operator $\hat \psi$ (see Eq. (\ref{II12})) 
which reproduces the operator $\hat Y_{SW}$ or $\hat Y_{C}$ accordingly to the choice $A=B$ or $A \neq B$.

iii) An algebraic study of the Shapiro-Wagner and Caves' schemes, performed in Section 4, shed a further light 
on the significant distinction between the two cases. The SW heterodyne is characterized by a symmetry expressed 
by a  noncompact subgroup of $SU(1,1)$ of the parabolic type \cite{LINDNAGEL}, while Caves' frame is characterized 
by a symmetry structure represented by a Lie  algebra constituted by a subalgebra of $su(1,1)$ type and by an 
Abelian maximal ideal. 

iv) An interpretation of Caves' heterodyning is described in Section 8.
Keeping in mind the RNS representation fairly working out in the Shapiro-Wagner case (Secs. 5,6 and 7), 
we define an operator $\hat D_{C}\equiv \hat S $ depending on the parameter $\mu =\sqrt{\frac{B}{A}}$  which 
constitutes an extension to ``noncommutative''  Caves' frame of the operator $\hat R $ used in the ``commutative'' 
RNS theory of the Shapiro-Wagner heterodyning.
Although our generalized formulae are given by exact expressions, they depend on the real parameter $\mu$ 
covering for $\mu=1$ the corresponding formulae of the SW frame.

Therefore,  the generalized formulae involving $\hat S$ found in Section 8 lend themselves to be elaborated 
{\it approximately} in the construction of an (approximate) RNS theory of quantum measurements of noncommuting 
variables (see Refs. \cite{Y1}, \cite{YL} and references therein).

\section{Quantum extension of the classical amplitude and phase concepts}

In Ref. \cite{NOVEL}, we have seen that the complex function
\begin{equation}
\label{HH1}
\psi (t)=\sqrt A e^{i\alpha}y_{1}(t) - \sqrt B e^{i\beta}y_{2}(t),
\end{equation}
$\alpha$, $\beta$, $A$, $B$ being real numbers, is important in getting the amplitude and the phase of a classical generalized oscillator, i.e. when the 
``mass'' and ``frequency'' may depend on time.
Below we shall report a concise summary of the theory developed in Ref. \cite{NOVEL}.
In (\ref{HH1}) $y_{1}(t)$ and $y_{2}(t)$ denote two independent solutions of the equation

\begin{equation}
\label{HH2}
\ddot y +\Omega^{2}(t) y=0,
\end{equation}
where $\Omega(t)$ is a given function of time.

Then, the function
\begin{equation}
\label{HH3}
\sigma = (Ay_{1}^{2} + By_{2}^{2} + 2Cy_{1}y_{2})^{\frac{1}{2}}
\end{equation}
satisfies the ordinary  nonlinear differential equation (known as the Ermakov-Milne-Pinney (EMP) equation) 
(see Refs. \cite{E},\cite{Mi},\cite{Pi},\cite{EG}). 
\begin{equation}
\label{HH4}
\ddot \sigma +\Omega^{2}(t)\sigma =\frac{1}{\sigma^{3}},
\end{equation}
with $A, B, C$ arbitrary constants such that
\begin{equation}
\label{HH5}
AB - C^{2}=\frac{1}{W_{0}^{2}}
\end{equation}
where $W_{0}$=$y_{1}\dot y_{2} - \dot y_{1} y_{2}$=${\it const}$ is the Wronskian. 

The time behavior of the phase $\theta _{cl}$ turns out to be expressed by 
\begin{equation}
\label{HH6}
\theta_{cl} (t)=\int^{t}_{t_{0}}\frac{dt'}{\sigma^{2}(t')}=F(t)-F(t_{0}),
\end{equation}
where
\begin{equation}
\label{HH7}
F(t)=\frac{1}{2i}{[\ln \psi (t) - \ln \psi^{*} (t)]}.
\end{equation}

On the other hand, the amplitude associated with $\theta_{cl}(t)$ is given by 
$\sigma_{cl}(t) \equiv \sigma (t)$ = $|\psi (t)|$, in the sense that

\begin{equation}
\label{HH8}
y(t)=\sigma _{cl}(t)(\theta_{cl}(t) + \delta)
\end{equation}
$(\delta=const)$. A simple check shows that, in the case of conventional oscillator 
$(\Omega \equiv \omega_{0} =const)$, we obtain $\dot F=\frac{1}{\sigma^{2}}=\omega_{0}$, 
$\theta_{cl}(t)=\omega_{0}(t-t_{0})$, as one expects.

What can we do about a possible quantum extension of this procedure?
In the following, we shall try to consider a simple canonical quantization of the abovementioned method 
limiting ourselves to the case of the conventional harmonic oscillator.
To this aim, we shall take into account the settling exploited, on one side, by Shapiro-Wagner \cite{SW} and, 
on another side, by Caves \cite{CAVES} in handling the quantum limits on simultaneous phase and 
squared-amplitude measurements established in optical heterodyne detection. In our frame both the scheme used 
by Shapiro-Wagner and by Caves can be dealt with in a compact and efficacious way, making deeply manifest the 
unifying potentiality of the method applied.

Really, the physical situation inherent to the SW frame is quite different from that relative to Caves' heterodyning, 
because in the first case the crucial operator realizing the apparatus is 
$\hat \psi_{SW}=\sqrt{A} \tilde a + \sqrt{B} \tilde b^{\dag}$ with $A=B$, 
where $[\hat \psi_{SW}, \hat \psi_{SW}^{\dag}]=0$. In contrast, in the second case, 
the operator realizing Caves' heterodyning is described by $\hat \psi_{C}=\sqrt{A} \tilde a + \sqrt{B}\tilde b^{\dag}$ 
with $A\neq B$, where $[\hat \psi_{C}, \hat \psi_{C}^{\dag}] \neq 0$ (see later).

An (ideal or feasible) {\it exact}  quantum phase can be built up for the SW frame {\it only}. 
However, for Caves' situation, in this work a formal approach is outlined (see Sec. 8) addressed to extend the 
RNS theory by means of the generalization of the operator $\hat R$ where $\hat \psi_{C}$, $\hat \psi_{C}^{\dag}$ 
do not commute.
Anyway, in Caves' framework phase and squared-amplitude cannot be defined in an exact manner. 
Thus, an approximate procedure turns out to be in order.
An undeniable advantage of our formalism is represented by the existence of {\it exact} formulae 
which are very indicated to be appropriately approximated (see Sec. 8).

At this point the mechanism and the concepts of the generalized quantum measurements reveal 
to be necessary tools to go on. This program can be pursued accordingly to the theory of generalized 
quantum measurements and approximate simultaneous measurements of noncommuting observables  expounded 
in Refs. \cite{Y1} and \cite{YL} and others (see references therein).
Quantization extension of the method adopted in Ref. \cite{NOVEL} to define the phase operator corresponding to 
the classical functions $\theta _{cl}$ and $F(t)$ (see Eqs. (\ref{HH6}) and (\ref{HH7})) can be carefully performed.
To be specific, what we think that it should be important in our formal interpretation of heterodyning is 
the operator structure of the quantum version of (classical) formula (\ref{HH1}), where to  the independent 
functions $y_{1}(t)$ and $y_{2}(t)$, it should correspond two independent operators $\hat a(t)$ and 
$\hat b^{\dag}(t)$, which should be interpreted as annihilation and creation operators of the signal 
and image fields, respectively. Another important aspect of our approach is the definition of an operator 
$\hat F (t)$ in such a way that it leads to an exact phase operator, in the SW case, corresponding to 
$ \theta_{cl}$, which is self adjoint, and such that a consistent Ban relative state phase representation be reliable.

In order to define appropriately an {\it exact} quantum phase and a squared-amplitude at least in the case of 
Shapiro-Wagner detector, and to set up an {\it approximate} theory of Caves' heterodyning on the basis of 
generalized quantum measurements framework, we believe that the quantization procedure could put aside the 
quantization of {\it all} the classical properties pertinent to the Ermakov-Milne-Pinney (EMP) equation.
In fact, while at the classical level one can define phase and amplitude of the field under consideration, 
following the properties of the EMP equation, at the quantum level we have two different physical situations.
In other words, in the SW case it is possible to define both a self-adjoint phase operator and an amplitude 
operator (\ref{Z16}), which could be interpreted as the quantized phase and amplitude of the field in the 
RNS representation. Vice versa, in Caves' frame, the phase operator cannot be {\it exactly} defined . 
Furthermore, the quantity which in the SW case is interpreted as an amplitude, is not well-defined in 
Caves' situation, since $[\hat \psi_{C}, \hat \psi_{C}^{\dag}] \neq 0$. Then, it turns out to be difficult 
to apply a correspondence principle which connects any classical quantity to a quantum operator.

To accomplish the preceding requirements, let us introduce the boson operators
\begin{equation}
\label{HH9}
\hat a(t) =\frac{\gamma^{*}}{\gamma}e^{i\omega_{0} t} \hat a_{0},
\end{equation}

\begin{equation}
\label{HH10}
\hat a(t)^{\dag} =\frac{\gamma}{\gamma^{*}}e^{-i\omega_{0} t} \hat a_{0}^{\dag},
\end{equation}

\begin{equation}
\label{HH11}
\hat b(t) =\frac{\gamma^{*}}{\gamma}e^{i\omega_{0} t} \hat b_{0},
\end{equation}

\begin{equation}
\label{HH12}
\hat b(t)^{\dag} =\frac{\gamma}{\gamma^{*}}e^{-i\omega_{0} t} \hat b_{0}^{\dag},
\end{equation}
so that (as it is customary)

\begin{equation}
\label{HH13}
[\hat a(t), \hat a^{\dag}(t)]=[\hat a_{0}, \hat a_{0}^{\dag}]=\hat 1,
\end{equation}
and

\begin{equation}
\label{HH14}
[\hat a(t), \hat b^{\dag}(t)]=[\hat a_{0}, \hat b_{0}^{\dag}]=0,
\end{equation}
where $\gamma=|\gamma|e^{i\delta_{0}}$ is a complex number.

The operators $\hat a(t)$ and $\hat b(t)$ would be interpreted as annihilation operators of a single-mode 
radiation field (signal) and of an image field, respectively.
To state a meaningful correspondence between the classical expression (\ref{HH1}) and a related operator 
coming from quantizing the field $\psi (t)$, we take the expectation values of $\hat a(t)$, $\hat b^{\dag}(t)$ 
on coherent states, so that

\begin{equation}
\label{HH15}
<\hat a(t)>=<\gamma|\frac{\gamma^{*}}{\gamma}e^{i\omega_{0}t} \hat a_{0}|\gamma>=
\gamma^{*}e^{i\omega_{0}t} \sim  e^{i\omega_{0}t} \rightarrow y_{1}(t),
\end{equation}
and

\begin{equation}
\label{HH16}
<\hat b^{\dag}(t)>=\gamma e^{-i\omega_{0}t} \sim e^{-i\omega_{0}t} \rightarrow y_{2}(t).
\end{equation}

We get also

\begin{equation}
\label{HH17}
<\hat a(t)><\hat b^{\dag}(t)>=|\gamma |^{2}.
\end{equation}

The mean values of $\hat a$ and $\hat b^{\dag}$ reproduce essentially the (classical) independent solutions 
$y_{1}(t)$ and $y_{2}(t)$. At the quantum level, $\hat a$ and $\hat b^{\dag}$ turn out to be independent 
(i.e. the signal and image fields) assuming that the total density operator $\rho$ factorizes, namely

\begin{equation}
\label{HH18}
\rho = \rho_{S}\otimes \rho _{I}
\end{equation}

Within the Shapiro-Wagner phase heterodyne detection, the operator (\ref{GG1}) is involved, whose properties 
are discussed in Ref. \cite{SW} (see also Refs. \cite{HRADIL} and \cite{B4}).
The  feasible phase in the framework of heterodyne detection was proposed by Shapiro-Wagner \cite{SW}. 
In this case, the measured quantity is considered a phase of the operator (\ref{GG1}). 
[$\hat a$, $\hat a^{\dag}$: annihilation and creation operators of a signal mode; $\hat b$, 
$\hat b^{\dag}$: annihilation and creation operators of an image band mode].
Hradil showed that the feasible phase is described by a unitary phase operator (the SW phase operator) 
defined by Eq. (\ref{GG8}) (Ref. \cite{HRADIL}). It should be noted that the SW phase operator is 
defined on the full Hilbert space 

\begin{equation}
\label{HH19}
\cal H = \cal H_{S}\otimes \cal H_{I},
\end{equation}
where $\cal H_{S}$, $\cal H_{I}$ are the Hilbert spaces for the signal and image modes.
To be precise, the short-hand notation (\ref{GG1}) is used by expressing that the heterodyne 
device described in SW realizes the quantum measurement

\begin{equation}
\label{HH20}
\hat Y_{SW}=\hat a \otimes \hat 1_{I} + \hat 1_{S} \otimes \hat b^{\dag}
\end{equation}
on the joint state space (\ref{HH19}), where $\hat 1_{I}$ and $\hat 1_{S}$ denote the identity operators 
on the spaces $\cal H_{I}$ and $\cal H_{S}$. The SW operator (\ref{HH20}) and its adjoint $\hat Y_{SW}^{\dag}$ commute.
Furthermore, $\hat D_{SW}$ obeys the commutation relation

\begin{equation}
\label{HH21}
[\hat D_{SW}, \hat N]=\hat D_{SW}
\end{equation}
($\hat N$ is the number-difference between the signal and local oscillator modes, i.e. $\hat N$=$\hat a^{\dag} \hat a$ - $\hat b^{\dag} \hat b$).

\section{A unified formulation of Shapiro-Wagner and Caves frames: the operators $\hat \psi$ and $\hat \psi^{\dag}$}

The quantum extension of the scalar field (\ref{HH1}) and its complex coniugate leads to the operators

\begin{equation}
\label{II1}
\hat \psi(t)=\sqrt{A} e^{i\alpha}\hat a (t)+\sqrt{B} e^{i\beta} \hat b^{\dag}(t),
\end{equation}

\begin{equation}
\label{II2}
\hat \psi^{\dag}(t)=\sqrt{A} e^{-i\alpha}\hat a^{\dag}(t)+\sqrt{B} e^{-i\beta} \hat b(t),
\end{equation}
where we have carried out the change $\beta \rightarrow \beta+\pi$ in order to obtain a quantity as soon 
as possible of the form $\hat a + \hat b^{\dag}$.

The operator (\ref{II1}) and (\ref{II2}) satisfy the commutation relation

\begin{equation}
\label{II3}
[\hat \psi, \hat \psi^{\dag}]=(A - B)\hat 1,
\end{equation}
from which two possible situations arise:

\noindent {\it (I)}
\begin{equation}
\label{II4}
[\hat \psi, \hat \psi^{\dag}]=0,
\end{equation}
for $A=B$, and

\noindent {\it II)}
\begin{equation}
\label{II5}
[\hat \psi, \hat \psi^{\dag}]\neq 0,
\end{equation}
for $A \neq B$. Equations (\ref{II4}) and (\ref{II5}) correspond to two distinct heterodyne devices. 
The first relies on the use of a photon detector in the heterodyne apparatus employed by Shapiro-Wagner, 
and the second refers to the behaviour of Caves' heterodyning square-law (power-detector).
A more detailed treatment of cases ${\it (I)}$ and ${\it (II)}$ will be shown in the following.
To this purpose let us put

\begin{equation}
\label{II6}
\hat a= \hat a_{1}+i\hat a_{2},\hat b= \hat b_{1}+i\hat b_{2},\hat a^{\dag}= \hat a_{1}-i\hat a_{2},\hat b^{\dag}= \hat b_{1}-i\hat b_{2},
\end{equation}
where $\hat a_{j} (j=1,2)$, $\hat b_{j} (j=1,2)$, are self-adjoint, but $\hat a, \hat b$ are not self-adjoint operators.
Now we build up the combinations
\begin{equation}
\label{II7}
\hat \psi + \hat \psi^{\dag}, \quad \hat \psi - \hat \psi^{\dag}.
\end{equation}

We easily have

\begin{equation}
\label{II8}
\hat y_{1}\equiv \frac{1}{2}(\hat \psi + \hat \psi^{\dag})=\sqrt{A}\tilde a_{1}+\sqrt{B}\tilde b_{1},
\end{equation}

\begin{equation}
\label{II9}
\hat y_{2}\equiv \frac{1}{2i}(\hat \psi - \hat \psi^{\dag})=\sqrt{A}\tilde a_{2}+\sqrt{B}\tilde b_{2},
\end{equation}
where $\hat y_{1}, \hat y_{2}$ are self-adjoint operators, and $\tilde a_{j} (j=1,2)$, $\tilde b_{j} (j=1,2)$ 
are defined as the components of the rotation-valued operators

\begin{equation}
\label{II10}
{\tilde a_{1}\choose \tilde a_{2}}=\left(\begin{array}{cc}
\cos\alpha & -\sin\alpha \\
\sin\alpha & \cos\alpha \\
\end{array} \right) {\hat a_{1}\choose \hat a_{2}},
\end{equation}

\begin{equation}
\label{II11}
{\tilde b_{1}\choose \tilde b_{2}}=\left(\begin{array}{cc}
\cos\beta & \sin\beta \\
-\sin\beta  & \cos\beta \\
\end{array} \right) {\hat b_{1}\choose \hat b_{2}},
\end{equation}
($ \beta $ stands for $\tilde \beta=\pi+\beta $).

Thus, the operators $\hat \psi$, $\hat \psi^{\dag}$ take the more concise forms

\begin{equation}
\label{II12}
\hat \psi = \sqrt{A}\tilde a + \sqrt{B}\tilde b^{\dag},
\end{equation}

\begin{equation}
\label{II13}
\hat \psi^{\dag} = \sqrt{A}\tilde a^{\dag} + \sqrt{B}\tilde b,
\end{equation}
where

\begin{equation}
\label{II14}
\tilde a=\tilde a_{1} + i\tilde a_{2}, \tilde b=\tilde b_{1}-i\tilde b_{2}.
\end{equation}

The requirement $[\tilde a, \tilde a^{\dag}]$ = $[\tilde b, \tilde b^{\dag}]$ = $\hat 1$, 
$[\tilde a, \tilde b]$=$[\tilde a^{\dag}, \tilde b^{\dag}]=0$, implies the commutation rule (\ref{II3}).

We point out that the Shapiro-Wagner frame is covered by Eq. (\ref{II12}) for $A=B$, so that in this case 
$\hat \psi$ can be identified by the operator $\hat  Y_{SW}=\tilde a_{S}+\tilde a_{I}^{\dag}$ in the 
rotation-valued operator representation, and the quadrature components of $\hat \psi$, $\hat y_{1}$ 
and $\hat y_{2}$, commute. 

On the other hand, for $A \neq B$, by choosing

\begin{equation}
\label{II15}
A=1+ \frac{\nu_{IF}}{\nu_{0}},  \qquad B=1- \frac{\nu_{IF}}{\nu_{0}},
\end{equation}
(see Refs. \cite{SW},\cite{CAVES},\cite{YUEN}) and inserting 
Eq. (\ref{II15}) into Eq. (\ref{II12}), we find 
an expression for the $\hat \psi$-operator for Caves' configuration in the rotation-valued representation, i.e.

\begin{equation}
\label{II16}
\hat \psi=(1+\frac{\nu_{IF}}{\nu_{0}})^{\frac{1}{2}}\tilde a + (1-\frac{\nu_{IF}}{\nu_{0}})^{\frac{1}{2}}\tilde b^{\dag}.
\end{equation}

In this case the quadrature operators $\hat y_{1}$, $\hat y_{2}$ given by Eqs. (\ref{II8}) and(\ref{II9}) do not commute, 
but provide just the formula
\begin{equation}
\label{II17}
[\hat y_{1}, \hat y_{2}]= \frac{1}{4i}[\hat \psi + \hat \psi^{\dag},\hat \psi - \hat \psi^{\dag}]=\frac{i}{2}(A-B)=i\frac{\nu_{IF}}{\nu_{0}}
\end{equation}
forseen in the case studied by Caves.

\section{Algebraic characterization of Shapiro-Wagner and Caves frameworks}

In the SW case, a simple Lie algebraic characterization of the heterodyne detection can be performed. 
To this aim we remind that the algebra of the noncompact group $SU(1,1)$ is defined by

\begin{equation}
\label{L1}
[J_{0},J_{1}]=iJ_{2}, [J_{0},J_{2}]=-iJ_{1}, [J_{1},J_{2}]=-iJ_{0},
\end{equation}
where

\begin{equation}
\label{L2}
\hat C = J_{0}^{2}-J_{1}^{2}-J_{2}^{2}
\end{equation}
is the Casimir invariant.

Sometimes one could use the ladder operators

\begin{equation}
\label{L3}
J_{\pm}=J_{1}\pm iJ_{\pm}
\end{equation}
satisfying the relations

\begin{equation}
\label{L4}
[J_{0},J_{\pm}]=\pm J_{\pm},
[J_{+},J_{-}]=-2J_{0}.
\end{equation}

Lindblad and Nagel \cite{LINDNAGEL} have shown that $SU(1,1)$ has three classes of conjugate 
one-parameter subgroups. The elliptic subgroups are compact, the hyperbolic and parabolic subgroups 
are noncompact. For example, the parabolic class is represented by the following specimen:

\begin{equation}
\label{L5}
n(\zeta)=
\left(\begin{array}{cc}
1-i\frac{\zeta}{2} & -i\frac{\zeta}{2} \\
i\frac{\zeta}{2} & 1+ i\frac{\zeta}{2} \\
\end{array} \right).
\end{equation}

The matrix $n(\zeta)$ is generated by

\begin{equation}
\label{L6}
K_{+}=J_{0}+J_{1}
\end{equation}
and

\begin{equation}
\label{L7}
[K_{+},J_{2}]=-iK_{+}
\end{equation}
holds.

A realization of the algebra (\ref{L1}) is given by

\begin{equation}
\label{L8}
J_{1}=\frac{1}{2}(\tilde a^{\dag} \tilde b^{\dag} + \tilde a \tilde b), J_{2}=
\frac{1}{2i}(\tilde a^{\dag} \tilde b^{\dag} - \tilde a \tilde b), J_{0}=
\frac{1}{2}(\tilde a^{\dag} \tilde a + \tilde b^{\dag} \tilde b +1).
\end{equation}

We have

\begin{equation}
\label{L9}
\frac{1}{2} \hat \psi^{\dag} \hat \psi =A(J_{0}+J_{1})\equiv K_{+}.
\end{equation}

In other words, when $[\hat \psi, \hat \psi^{\dag}]=0$ $(A\equiv B)$ the operator 
$\frac{1}{2} \hat \psi^{\dag} \hat \psi$ realizes the {\it parabolic} $SU(1,1)$ subgroup. 

The Casimir invariant can be easily evaluated. We obtain

\begin{equation}
\label{L10}
\hat C=\frac{1}{4}(\hat N^{2}-\hat 1)
\end{equation}
$(\hat N=\tilde a^{\dag}\tilde a - \tilde b^{\dag} \tilde b)$.

In Caves' framework, $[\hat \psi, \hat \psi^{\dag}]=(A-B)\hat 1 $ $\neq 0$.
The operators $\hat \psi \hat \psi^{\dag}$, $\hat \psi^{\dag} \hat \psi $ take the form

\begin{equation}
\label{L11}
\hat \psi \hat \psi^{\dag}=A \tilde a^{\dag} \tilde a + B \tilde b^{\dag} \tilde b + 
\sqrt{AB}(\tilde a \tilde b + \tilde a^{\dag} \tilde b^{\dag}) + A,
\end{equation}

\begin{equation}
\label{L12}
\hat \psi^{\dag} \hat \psi^=A \tilde a^{\dag} \tilde a + B \tilde b^{\dag} \tilde b + \sqrt{AB}(\tilde a \tilde b + \tilde a^{\dag} \tilde b^{\dag}) + B.
\end{equation}

It is convenient to introduce the operators

\begin{equation}
\label{L13}
N_{1}=\frac{1}{2}(\tilde a^{\dag} \tilde a + \tilde b^{\dag} \tilde b +1),
\end{equation}

\begin{equation}
\label{L14}
N_{2}=\frac{1}{2}(\tilde a^{\dag} \tilde a -\tilde b^{\dag} \tilde b -1),
\end{equation}
so that

\begin{equation}
\label{L15}
N_{1}+N_{2}= \tilde a^{\dag} \tilde a
\end{equation}

\begin{equation}
\label{L16}
N_{1}-N_{2}= \tilde b^{\dag} \tilde b + 1,
\end{equation}

\begin{equation}
\label{L17}
L_{1}=\frac{1}{2}(\tilde a^{\dag} \tilde b^{\dag} +\tilde a \tilde b),L_{2}=\frac{1}{2i}(\tilde a^{\dag} \tilde b^{\dag} -\tilde a \tilde b),
\end{equation}

\begin{equation}
\label{L18}
L_{+}=L_{1}+iL_{2}=\tilde a^{\dag} \tilde b^{\dag}, L_{-}=L_{1}-iL_{2}=\tilde a\tilde b.
\end{equation}

Taking account of the commutation relations

\begin{equation}
\label{L19}
[\tilde a^{\dag} \tilde a, \tilde b^{\dag} \tilde b]=0, 
[\tilde a^{\dag} \tilde a,\tilde a \tilde b]=-\tilde a \tilde b, 
 [\tilde a^{\dag} \tilde a, \tilde a^{\dag} \tilde b^{\dag}]=\tilde a^{\dag} \tilde b^{\dag},
\end{equation}

\begin{equation}
\label{L20}
[\tilde b^{\dag} \tilde b, \tilde a \tilde b]=-\tilde a \tilde b, 
[\tilde b^{\dag} \tilde b, \tilde a^{\dag} \tilde b^{\dag}]=\tilde a^{\dag} \tilde b^{\dag}, 
[\tilde a \tilde b, \tilde a^{\dag} \tilde b^{\dag}]=\tilde a^{\dag} \tilde a + \tilde b^{\dag} \tilde b + 1,
\end{equation}
we arrive at the algebra described by the commutation relations

\begin{equation}
\label{L21}
[L_{+},L_{-}]=-2N_{1},
\end{equation}

\begin{equation}
\label{L22}
[L_{+}, N_{1}]=-L_{+},
\end{equation}

\begin{equation}
\label{L23}
[L_{-}, N_{1}]=L_{-},
\end{equation}

\begin{equation}
\label{L24}
[L_{+},N_{2}]=0,
\end{equation}

\begin{equation}
\label{L25}
[L_{-},N_{2}]=0,
\end{equation}

\begin{equation}
\label{L26}
[N_{1},N_{2}]=0,
\end{equation}
where

\begin{equation}
\label{L27}
L_{+}=\tilde a^{\dag} \tilde b^{\dag},
\end{equation}

\begin{equation}
\label{L28}
L_{-}=\tilde a \tilde b,
\end{equation}

\begin{equation}
\label{L29}
N_{1}=\frac{1}{2}(\tilde a^{\dag} \tilde a + \tilde b^{\dag} \tilde b +1),
\end{equation}

\begin{equation}
\label{L30}
N_{2}=\frac{1}{2}(\tilde a^{\dag} \tilde a - \tilde b^{\dag} \tilde b -1).
\end{equation}

The algebra (\ref{L21})-(\ref{L26}) refers to the Caves microwave heterodyning with square-law detectors.
The algebra constituted by (\ref{L21})-(\ref{L23}) is of the $su(1,1)$ type realized by

\begin{equation}
\label{L31}
L_{+}=\tilde a^{\dag} \tilde b^{\dag},L_{-}=\tilde a \tilde b, N_{1}=
\frac{1}{2}(\tilde a^{\dag} \tilde a + \tilde b^{\dag} \tilde b + 1),
\end{equation}
while the Abelian commutation rules (\ref{L24})-(\ref{L26}) define the maximal ideal 
(or the centre) of the algebra of elements (\ref{L27})-(\ref{L30}).
In Caves' configuration, the expressions for $\hat \psi \hat \psi^{\dag}$ and $\hat \psi^{\dag} \hat \psi$ read

\begin{equation}
\label{L32}
\hat \psi \hat \psi^{\dag}=A(N_{1}+N_{2}+1)+B(N_{1}-N_{2}-1)+\sqrt{AB}(L_{+}+L_{-}),
\end{equation}

\begin{equation}
\label{L33}
\hat \psi^{\dag} \hat \psi= A(N_{1}+N_{2})+B(N_{1}-N_{2})+\sqrt{AB}(L_{-}+L_{+}).
\end{equation}

If $A=B$ the Shapiro-Wagner scheme

\begin{equation}
\label{L34}
\frac{1}{2}\hat \psi^{\dag}\hat \psi=\frac{1}{2}\hat \psi \hat \psi^{\dag}=A(J_{0}+J_{1})\equiv AK_{+}
\end{equation}
arises (Refs. \cite{B1}-\cite{B3}).

\section{The relative number state representation of the phase operator:  the feasible Shapiro-Wagner and RNS phase operators}

An appropriate definition of a phase quantum mechanical operator presents notable difficulties 
[see, for instance, the review by Carruthers-Nieto and Lynch for the description of the main aspects 
of this long story \cite{CN},\cite{L}],

A remarkable wayout to a more consistent definition of a phase operator and its  eigenstates is obtained by 
the socalled relative number state (RNS) representation, proposed by 
Ban \cite{B1}-\cite{B3}.
In our approach, the operator (\ref{II12}) (and its adjoint $\hat \psi^{\dag}$), in the case in which $A=B$, 
can be identified by the Shapiro-Wagner complex amplitude (in the rotation-valued operator representation) 
(and its adjoint) whose phase is referred as the {\it feasible} phase within the framework of heterodyne 
detection proposed in Ref. \cite{SW}. As we have recalled previously, two ways can be settled up to define 
quantum mechanical phase operators. One is based on the polar decomposition of the annihilation operator 
of a photon ({\it ideal} phase), and the other on the use of phase-measurement processes ({\it feasible} phase).
[See, for example, Ref. \cite{B4} and references therein]. In our formalism, we can write

\begin{equation}
\label{N1}
\hat \psi \equiv \hat Y_{SW}=\sqrt{A}(\tilde a +\tilde b^{\dag})
\end{equation}
where $\tilde a, \tilde a^{\dag}$ are annihilation and creation operators of a signal mode, 
and $\tilde b, \tilde b^{\dag}$ are those of an image band mode (in the rotation-valued 
representation defined by Eq. (\ref{II14}) and Eqs. (\ref{II10})-(\ref{II11})).

The {\it  feasible} phase introduced by Shapiro-Wagner (SW phase operator) 
is given in our approach by the {\it unitary operator}
\begin{equation}
\label{N2}
D_{SW}=\hat \psi(\hat \psi^{\dag} \hat \psi)^{-\frac{1}{2}},
\end{equation}
which is defined on the extended Hilbert space $\cal H=\cal H_{A}\otimes \cal H_{B}$, where $\cal H_{A}$ is the 
Hilbert space of the signal photon $(\tilde a, \tilde a^{\dag})$ and $\cal H_{B}$ is the Hilbert space of the 
mode described by $(\tilde b, \tilde b^{\dag})$ (see \cite{B4}).

Since in the SW scheme $[\hat \psi, \hat \psi^{\dag}]=0$, Eq. (\ref{N2}) provides

\begin{equation}
\label{N3}
D_{SW}\equiv \hat R= \hat \psi^{\frac{1}{2}} (\hat \psi^{\dag})^{-\frac{1}{2}}=
\sqrt{\frac{\tilde a + \tilde b^{\dag}}{\tilde a^{\dag} + \tilde b}}.
\end{equation}

The unitary operator $\hat R$ obeys the commutation rule

\begin{equation}
\label{N4}
[\hat R, \hat N]=\hat R.
\end{equation}

Relation (\ref{N4}) can be proved by resorting to the properties \cite{LOUISELL}

\begin{equation}
\label{N5}
[a, f(a,a^{\dag})]=\frac{\partial f}{\partial a^{\dag}},
\end{equation}

\begin{equation}
\label{H6}
[a^{\dag}, f(a,a^{\dag})]=-\frac{\partial f}{\partial a}.
\end{equation}

In fact, applying Eqs. (\ref{N5}) and (\ref{H6}) we deduce

\begin{equation}
\label{N7}
[\tilde a, (\tilde a^{\dag} + \tilde b)^{-\frac{1}{2}}]=
\frac{\partial (\tilde a^{\dag} + \tilde b)^{-\frac{1}{2}}}{\partial \tilde a ^{\dag}}=
-\frac{1}{2}(\tilde a^{\dag} +\tilde b)^{-\frac{3}{2}},
\end{equation}

\begin{equation}
\label{N8}
[\tilde a^{\dag}, (\tilde a + \tilde b^{\dag})^{\frac{1}{2}}]=
-\frac{\partial (\tilde a + \tilde b^{\dag})^{\frac{1}{2}}}{\partial \tilde a}=-\frac{1}{2}(\tilde a + \tilde b^{\dag})^{-\frac{1}{2}},
\end{equation}

\begin{equation}
\label{N9}
[\tilde a, (\tilde a + \tilde b^{\dag})^{\frac{1}{2}}]=
[\tilde a^{\dag}, (\tilde a^{\dag} + \tilde b)^{-\frac{1}{2}}]=0,
\end{equation}

\begin{equation}
\label{N10}
\tilde a (\tilde a^{\dag}+ \tilde b)^{-\frac{1}{2}}=
(\tilde a^{\dag}+ \tilde b)^{-\frac{1}{2}}\tilde a - \frac{1}{2}(\tilde a^{\dag} + \tilde b)^{-\frac{3}{2}},\end{equation}

\begin{equation}
\label{N11}
\tilde b (\tilde a + \tilde b^{\dag})^{\frac{1}{2}}=
(\tilde a + \tilde b^{\dag})^{\frac{1}{2}} \tilde b + \frac{1}{2}(\tilde a + \tilde b^{\dag})^{-\frac{1}{2}}.
\end{equation}

Taking account of Eqs. (\ref{N7})-(\ref{N11}), the commutation rule (\ref{N4}) is readily determined.

\section{Evaluation of the commutator $[\hat \theta, \hat N]$}

We have seen that, at the classical level, the function

\begin{equation}
\label{M1}
F(t)=\frac{1}{2i}[\ln \psi(t) - \ln \psi^{*}(t)]
\end{equation}
furnishes the (classical) phase

\begin{equation}
\label{M2}
\theta _{cl}(t)=F(t) - F(t_{0}).
\end{equation}

The quantum extension of this procedure yields

\begin{equation}
\label{M3}
\theta _{cl} \rightarrow \hat \theta = \hat F(t) - \hat F (t_{0})=
\frac{1}{2i}[\ln \hat \psi (t) - \ln \hat \psi^{\dag}(t)] - \frac{1}{2i}[\ln \hat \psi_{0} - \ln \hat \psi^{\dag}_{0}],
\end{equation}
where $\hat F(t_{0})$ is an arbitrary constant operator.
Assuming the SW scheme ($[\hat \psi,\hat \psi^{\dag}]=0$ ($A=B$)), Eq. (\ref{M3}) can be written as

\begin{equation}
\label{M4}
\hat \theta=\frac{1}{2i}[\ln(\sqrt{A}(\tilde a + \tilde b^{\dag}))-\ln(\sqrt{A}(\tilde a^{\dag} + \tilde b))]-
\frac{1}{2i}[\ln(\sqrt{A}(\tilde a_{0}+\tilde b^{\dag}_{0})) - \ln (\sqrt{A}(\tilde a^{\dag}_{0} + \tilde b_{0}))],
\end{equation}
namely

\begin{equation}
\label{M5}
\hat \theta=\frac{1}{2i}[\ln(\tilde a + \tilde b^{\dag})-\ln(\tilde a^{\dag} + \tilde b)]-
\frac{1}{2i}[\ln(\tilde a_{0}+\tilde b^{\dag}_{0}) - \ln (\tilde a^{\dag}_{0} + \tilde b_{0})],
\end{equation}
in the rotation-valued operator representation $(\tilde a, \tilde b)$, $(\tilde a^{\dag}, \tilde b^{\dag})$.

Up to a constant operator $\hat F (t_{0})$, we obtain

\begin{equation}
\label{M6}
\hat \theta = \frac{1}{2i}[\ln (\tilde a + \tilde b^{\dag})- \ln (\tilde a^{\dag} + \tilde b)].
\end{equation}

Since $\hat \theta^{\dag}=\hat \theta$, the operator (\ref{M6}) is self-adjoint. 
Now it is instructive to calculate the commutator $[\hat \theta, \hat N]$.
To this aim, some preliminary statements are in order.
We notice that from (\ref{N3})

\begin{equation}
\label{M7}
\ln \hat R=\ln \sqrt{\frac{\tilde a + \tilde b^{\dag}}{\tilde a^{\dag} + \tilde b}},
\end{equation}
which entails 
\begin{equation}
\label{M8}
\frac{1}{i}\ln \hat R =\frac{1}{2i}[\ln (\tilde a + \tilde b^{\dag})-\ln (\tilde a^{\dag} + \tilde b)]\equiv \hat \theta,
\end{equation}
viz.

\begin{equation}
\label{M9}
\hat R = e^{i\hat \theta}.
\end{equation}

We have

\begin{equation}
\label{M10}
\hat R = e^{i\hat \theta} = \sqrt{\frac{\tilde a + \tilde b^{\dag}}{\tilde a^{\dag} + \tilde b}},
\end{equation}

\begin{equation}
\label{M11}
\hat R^{\dag} = e^{-i\hat \theta^{\dag}} = \sqrt{\frac{\tilde a^{\dag} + \tilde b}{\tilde a + \tilde b^{\dag}}},
\end{equation}
where $\hat \theta=\hat \theta^{\dag}$.

It is a simple matter to check that $\hat R \hat R^{\dag} = \hat R^{\dag} \hat R = \hat 1$, i.e. $\hat R$ is unitary.

Since $\hat \theta $ is self-adjoint, it is possible to define the operators

\begin{equation}
\label{M12}
\hat \cos \theta = \frac{1}{2}(e^{i\hat \theta}+e^{-i\hat \theta})=\frac{1}{2}(\hat R + \hat R^{\dag}),
\end{equation}

\begin{equation}
\label{M13}
\hat \sin \theta = \frac{1}{2i}(e^{i\hat \theta}-e^{-i\hat \theta})=\frac{1}{2i}(\hat R - \hat R^{\dag}).
\end{equation}

The following properties

\begin{equation}
\label{M14}
[\hat \cos \theta,\hat \sin \theta] = -\frac{1}{2i}[\hat R, \hat R^{\dag}]=-\frac{1}{2i}[e^{i\hat \theta}, e^{-i\hat \theta}]=0,
\end{equation}

\begin{equation}
\label{M15}
\hat \cos^{2} \theta + \hat \sin^{2} \theta =\frac{1}{2}\{\hat R, \hat R^{\dag} \} \equiv \hat 1,
\end{equation}

\begin{equation}
\label{M16}
\hat \cos^{2} \theta - \hat \sin^{2} \theta =\frac{1}{2}(\hat R^{2}+ (\hat R^{2})^{\dag})
\end{equation}

hold.

Below we shall prove the commutation relation

\begin{equation}
\label{M17}
[\hat \theta, \hat N]=-i,
\end{equation}
where

\begin{equation}
\label{M18}
\hat \theta = \frac{1}{2i}[\ln \hat \psi - \ln \hat \psi^{\dag}] \equiv \hat \theta^{\dag}
\end{equation}

\begin{equation}
\label{M19}
\hat N=\tilde a^{\dag} \tilde a - \tilde b^{\dag} \tilde b
\end{equation}

\begin{equation}
\label{M20}
\hat \psi =\sqrt{A}(\tilde a + \tilde b^{\dag})
\end{equation}

\begin{equation}
\label{M21}
\hat \psi^{\dag} =\sqrt{A}(\tilde a^{\dag} + \tilde b).
\end{equation}

By using the relations (\ref{N5})-(\ref{H6}) (Ref. \cite{LOUISELL}) we obtain

\begin{equation}
\label{M22}
[\tilde a, \ln(\tilde a + \tilde b^{\dag})]=\frac{\partial \ln (\tilde a + \tilde b^{\dag})}{\partial \tilde a^{\dag}}=0 
\end{equation}

\begin{equation}
\label{M23}
[\tilde a^{\dag}, \ln(\tilde a + \tilde b^{\dag})]=
-\frac{\partial \ln (\tilde a + \tilde b^{\dag})}{\partial \tilde a}=-\frac{1}{\tilde a + \tilde b^{\dag}}
\end{equation}

\begin{equation}
\label{M24}
[\tilde a, \ln(\tilde a^{\dag} + \tilde b)]=
\frac{\partial \ln (\tilde a^{\dag} + \tilde b)}{\partial \tilde a^{\dag}}=\frac{1}{\tilde a^{\dag} + \tilde b}
\end{equation}

\begin{equation}
\label{M25}
[\tilde a^{\dag}, \ln(\tilde a^{\dag} + \tilde b)]=-\frac{\partial \ln (\tilde a^{\dag} + \tilde b)}{\partial \tilde a}=0
\end{equation}

\begin{equation}
\label{M26}
[\tilde b, \ln(\tilde a + \tilde b^{\dag})]=
\frac{\partial \ln (\tilde a + \tilde b^{\dag})}{\partial \tilde b^{\dag}}=\frac{1}{\tilde a + \tilde b^{\dag}}
\end{equation}

\begin{equation}
\label{M27}
[\tilde b^{\dag}, \ln(\tilde a + \tilde b^{\dag})]=
-\frac{\partial \ln (\tilde a + \tilde b^{\dag})}{\partial \tilde b}=0
\end{equation}

\begin{equation}
\label{M28}
[\tilde b, \ln(\tilde a^{\dag} + \tilde b)]=\frac{\partial \ln (\tilde a^{\dag} + \tilde b)}{\partial \tilde b^{\dag}}=0
\end{equation}

\begin{equation}
\label{M29}
[\tilde b^{\dag}, \ln(\tilde a^{\dag} + \tilde b)]=
-\frac{\partial \ln (\tilde a^{\dag} + \tilde b)}{\partial \tilde b}=-\frac{1}{\tilde a^{\dag} + \tilde b}.
\end{equation}

Then

\begin{equation}
\label{M30}
[\ln \hat \psi, \tilde a^{\dag} \tilde a ]=\frac{\tilde a}{\tilde a + \tilde b^{\dag}}
\end{equation}

\begin{equation}
\label{M31}
[\ln \hat \psi^{\dag}, \tilde a^{\dag} \tilde a ]=-\frac{\tilde a^{\dag}}{\tilde a^{\dag} + \tilde b}
\end{equation}

\begin{equation}
\label{M32}
[\ln \hat \psi, \tilde b^{\dag} \tilde b ]=-\frac{\tilde b^{\dag}}{\tilde a + \tilde b^{\dag}}
\end{equation}

\begin{equation}
\label{M33}
[\ln \hat \psi^{\dag}, \tilde b^{\dag} \tilde b ]=\frac{\tilde b}{\tilde a^{\dag} + \tilde b}.
\end{equation}

Inserting Eqs. (\ref{M30})-(\ref{M33}) into the equation

\begin{equation}
\label{M34}
[\hat \theta , \hat N]= 
\frac{1}{2i}[\ln \hat \psi - \ln \hat \psi^{\dag}, \tilde a^{\dag} \tilde a - \tilde b^{\dag} \tilde b ]
\end{equation}
we find

\begin{equation}
\label{M35}
[\hat \theta , \hat N]= \frac{1}{2i}
\left( \frac{\tilde a}{\tilde a + \tilde b^{\dag}}+ \frac{\tilde a^{\dag}}{\tilde a^{\dag} + \tilde b} 
+ \frac{\tilde b^{\dag}}{\tilde a + \tilde b^{\dag}} + \frac{\tilde b}{ \tilde a^{\dag} + \tilde b} \right)= -i.
\end{equation}

\section{Quadrature components of $\hat \psi$ in the Shapiro-Wagner scheme}

Let us define the self-adjoint operators (which can be interpreted as the {\it quadrature components} of $\hat \psi = \sqrt{A}(\tilde a + \tilde b^{\dag})$ in the SW heterodyne detector):

\begin{equation}
\label{Z1}
\hat y_{1} = \frac{1}{2}(\hat \psi + \hat \psi^{\dag})=\frac{\sqrt{A}}{2}[(\tilde a + \tilde b^{\dag})+(\tilde a^{\dag} + \tilde b)]
\end{equation}

\begin{equation}
\label{Z2}
\hat y_{2} =\frac{1}{2i}(\hat \psi - \hat \psi^{\dag})=\frac{\sqrt{A}}{2i}[(\tilde a +\tilde b^{\dag}) - (\tilde a^{\dag} + \tilde b)].
\end{equation}

We have

\begin{equation}
\label{Z3}
\hat y_{1}^{2}+ \hat y_{2}^{2}=A[\tilde a^{\dag} \tilde a +\tilde b^{\dag}\tilde b + 1 +(\tilde a \tilde b + \tilde a^{\dag} \tilde b^{\dag})]=2A(J_{0}+J_{1}),
\end{equation}
with

\begin{equation}
\label{Z4}
J_{1}=\frac{1}{2}(\tilde a^{\dag}\tilde b^{\dag} + \tilde a \tilde b), J_{2}=\frac{1}{2i}(\tilde a^{\dag} \tilde b^{\dag} - \tilde a \tilde b), J_{0}=\frac{1}{2}(\tilde a^{\dag} \tilde a + \tilde b^{\dag} \tilde b +1)
\end{equation}
(self-adjoint operators).

\subsection{Further elaborations}

Previously we have seen that (in the SW scheme)

\begin{equation}
\label{ZR1}
\hat R = e^{i\hat \theta} = \sqrt \frac{\tilde a + \tilde b^{\dag}}{\tilde a^{\dag} + \tilde b},
\end{equation}

\begin{equation}
\label{ZR2}
\hat R^{\dag} = e^{-i\hat \theta} = \sqrt \frac{\tilde a^{\dag} + \tilde b}{\tilde a + \tilde b^{\dag}},
\end{equation}
so that

\begin{equation}
\label{Z5}
\hat \cos \theta =\frac{1}{2}(\hat R + \hat R^{\dag})=
\frac{1}{2} \left[ \sqrt\frac{\tilde a + \tilde b^{\dag}}{\tilde a^{\dag} + \tilde b} + 
\sqrt\frac{\tilde a^{\dag} + \tilde b}{\tilde a + \tilde b^{\dag}} \right]
\end{equation}
 
\begin{equation}
\label{Z6}
\hat \sin \theta =\frac{1}{2i}(\hat R - \hat R^{\dag})=
\frac{1}{2i} \left[ \sqrt\frac{\tilde a + \tilde b^{\dag}}{\tilde a^{\dag} + \tilde b} - 
\sqrt\frac{\tilde a^{\dag} + \tilde b}{\tilde a + \tilde b^{\dag}} \right].
\end{equation}
 
Since

\begin{equation}
\label{Z7}
\hat \psi =\sqrt{A}(\tilde a + \tilde b^{\dag})
\end{equation}

\begin{equation}
\label{Z71}
\hat \psi^{\dag} =\sqrt{A}(\tilde a^{\dag} + \tilde b),
\end{equation}
we find

\begin{equation}
\label{Z8}
\sqrt{\tilde a + \tilde b^{\dag}}=\frac{\hat \psi^{\frac{1}{2}}}{A^{\frac{1}{4}}}
\end{equation}

\begin{equation}
\label{Z9}
\sqrt{\tilde a^{\dag} + \tilde b}=\frac{(\hat \psi^{\frac{1}{2}})^{\dag}}{A^{\frac{1}{4}}}
\end{equation}

The operators $\hat \cos \theta$, $ \hat \sin \theta $ can be expressed in terms of $\hat \psi$, $\hat \psi^{\dag}$, namely

\begin{equation}
\label{Z10}
\hat \cos \theta =\frac{1}{2}[\hat \psi \hat \psi^{\dag}]^{\frac{1}{2}} (\hat \psi + \hat \psi^{\dag}),
\end{equation}

\begin{equation}
\label{Z11}
\hat \sin \theta =\frac{1}{2i}[\hat \psi \hat \psi^{\dag}]^{\frac{1}{2}} (\hat \psi - \hat \psi^{\dag}).
\end{equation}

An easy check gives $\hat \cos^{2}\theta +  \hat \sin^{2}\theta =\hat 1$.

Equations (\ref{Z10}), (\ref{Z11}) take the form

\begin{equation}
\label{Z12}
\hat \cos \theta = \frac{\hat y_{1}}{[\hat \psi \hat \psi^{\dag}]^{\frac{1}{2}}},\end{equation}

\begin{equation}
\label{Z13}
\hat \sin \theta = \frac{\hat y_{2}}{[\hat \psi \hat \psi^{\dag}]^{\frac{1}{2}}},\end{equation}
so that the quadrature operators $\hat y_{1}$, $\hat y_{2}$ in the SW case can be represented by

\begin{equation}
\label{Z14}
\hat y_{1}= [\hat \psi \hat \psi^{\dag}]^{\frac{1}{2}}\hat \cos \theta,
\end{equation}

\begin{equation}
\label{Z15}
\hat y_{2}= [\hat \psi \hat \psi^{\dag}]^{\frac{1}{2}}\hat \sin \theta.
\end{equation}

Hence,

\begin{equation}
\label{Z16}
\hat y_{1}^{2}+ \hat y_{2}^{2}=\hat \psi \hat \psi^{\dag}=\hat \psi^{\dag} \hat \psi \equiv \hat \Lambda^{2},
\end{equation}
where $\hat \Lambda$ can be regarded as the SW amplitude operator.

\subsection{An interesting property of the quadrature operators $\hat y_{1}$, $\hat y_{2}$}

By using the property

\begin{equation}
\label{Z0}
\hat \psi^{\dag} \hat \psi = 2A(J_{0}+J_{1})
\end{equation}
(see Eq. (\ref{L34})) we get the representation

\begin{equation}
\label{Z17}
\hat y_{1} = \sqrt{2A(J_{0}+J_1)} \hat \cos \theta,
\end{equation}

\begin{equation}
\label{Z18}
\hat y_{2} = \sqrt{2A(J_{0}+J_1)} \hat \sin \theta,
\end{equation}
of the quadrature operators $\hat y_{1}$, $\hat y_{2}$ for the SW heterodyne detection.

Equations (\ref{Z14}) and (\ref{Z15}) yield Eq. (\ref{Z16}), in which the sum of squared-quadrature 
operators can be interpreted as quantities proportional to the parabolic $SU(1,1)$ subgroup generator.

\section{Interpretation of Caves' heterodyne detection}

To deal with the theory of amplitude and phase detection, it is convenient to introduce the operators

\begin{equation}
\label{C1}
\hat D_{C} \equiv \hat S = \frac{1}{\sqrt{2}}\sqrt{\hat T \hat Z^{-1} + \hat Z^{-1}\hat T},
\end{equation}
and

\begin{equation}
\label{C2}
\hat D_{C}^{\dag} \equiv \hat S^{\dag}=
\frac{1}{\sqrt{2}}\sqrt{ (\hat Z^{\dag})^{-1}\hat T^{\dag} + \hat T^{\dag} (\hat Z^{-1})^{\dag}},
\end{equation}
where

\begin{equation}
\label{C3}
\hat T = \tilde a + \mu \tilde b^{\dag}, \hat Z = \tilde a^{\dag} + \mu \tilde b
\end{equation}
and

\begin{equation}
\label{C4}
\mu = \sqrt{\frac{B}{A}} \neq 1.
\end{equation}

The definition (\ref{C1}) is a consequence of the fact that now $\hat T$ and $\hat Z $ do not commute:

\begin{equation}
\label{C5}
[\hat T, \hat Z]= (1- \mu^{2})\hat 1=\frac{A-B}{A} \hat 1 \equiv k \hat 1 \neq 0.
\end{equation}

The operators (\ref{C3}) are linked to $\hat \psi_{C}$ and $(\hat \psi_{C})^{\dag}$ in the sense that

\begin{equation}
\label{C6}
\hat \psi_{C}=\sqrt{A}\hat T,
\end{equation}

\begin{equation}
\label{C7}
(\hat \psi_{C})^{\dag} = \sqrt{A}\hat Z,
\end{equation}
with

\begin{equation}
\label{C8}
[\hat \psi_{C}, (\hat \psi_{C})^{\dag}]=(A - B)\hat 1.
\end{equation}

To study the operators $\hat S$, $\hat S^{\dag}$ in relation to a possible unitarity property, let us 
build up the products $\hat S \hat S^{\dag}$ and $\hat S^{\dag} \hat S$. After some manipulations, we find

\begin{equation}
\label{C9}
\hat S \hat S^{\dag}=\frac{1}{2}\sqrt{4 - k(\hat T \hat Z)^{-1} + k(\hat T \hat Z - k)^{-1}},
\end{equation}

\begin{equation}
\label{C10}
\hat S^{\dag} \hat S^=\frac{1}{2}\sqrt{4 - k(\hat T \hat Z)^{-1} + k(\hat T \hat Z + k)^{-1}}.
\end{equation}

The ingredients used to derive (\ref{C9}) and (\ref{C10}) are the commutation rules (\ref{C5}) and

\begin{equation}
\label{C11}
[\hat Z^{-1}, \hat T]= k\hat Z^{-2},
\end{equation}

\begin{equation}
\label{C12}
[\hat Z, \hat T^{-1}]= k\hat T^{-2},
\end{equation}
where the property $\hat T^{\dag} = \hat Z$ has been considered.

We observe that $\hat S \hat S^{\dag}\neq \hat S^{\dag} \hat S \neq \hat 1$ (for $k \neq 0$). 
Thus, in Caves' scheme the operator (\ref{C1}) can {\it not} be {\it exactly} unitary. However, 
$\hat S$ can be considered ``unitary'' in an {\it approximate} sense ($k \sim 0$). 
This is meaningful by a physical point of view. In fact, for example, under the hypothesis 
$\nu_{IF}<<\nu_{0}$ (invoked in Ref. \cite{SW} also), we have (see Eq. (\ref{II15}))
\begin{equation}
\label{C13}
k=\frac{A-B}{A}=2\frac{\nu_{IF}}{\nu_{0}}(1+\frac{\nu_{IF}}{\nu_{0}})^{-1}\sim 2\frac{\nu_{IF}}{\nu_{0}}(1-\frac{\nu_{IF}}{\nu_{0}})
\end{equation}
at the first order in $\frac{\nu_{IF}}{\nu_{0}}$.

So the methodological prescription to treat Caves' configuration heterodyning is that outlined by Yuen \cite{Y1}-\cite{YL} of 
generalized quantum measurements of noncommuting variables (see the references quoted in Refs. \cite{Y1} and \cite{YL} as well).
An important feature of the operators (\ref{C9}) and (\ref{C10}) is that these cover the Shapiro-Wagner and Caves' scheme.
To dwell upon this problem, the operators (\ref{C1}) and (\ref{C2}) should play a fundamental role in a relative number state 
representation (RNS) theory of quantum phase. 
Of course, according to Ref. \cite{Y1} generalized quantum measurements could 
be involved. These should be formally realizable as {\it approximate} simultaneous measurements of noncommuting observables.

\subsection{Relative number state representation approach of Caves' heterodyning}

In analogy with the RNS representation theory of phase operator of the Shapiro-Wagner type based on 
the operator $\hat R$ studied in Sections 5 and 7, here we would explore a possible RNS approach of 
Caves' scheme on the ground of the operators $\hat S$ and $\hat S^{\dag}$ defined by Eqs. (\ref{C1}) and (\ref{C2}).
To this aim, for a better understanding of the main differences between the ``commutative'' and the ``noncommutative'' 
frameworks of detections, it is convenient to express $\hat S$ and $\hat S^{\dag}$ in terms of the operators 
$\hat \psi_{C}$ and $ (\hat \psi_{C})^{\dag}$ (see Eqs. (\ref{C6}) and (\ref{C7})).

It results

\begin{equation}
\label{C14}
\hat S=\frac{1}{\sqrt{2}}\sqrt{\hat \psi_{C} (\hat \psi_{C}^{\dag})^{-1}+(\hat \psi_{C}^{\dag})^{-1} \hat \psi_{C}},
\end{equation}

\begin{equation}
\label{C15}
\hat S^{\dag}=\frac{1}{\sqrt{2}}\sqrt{(\hat \psi_{C})^{-1}\hat \psi_{C}^{\dag}+\hat \psi_{C}^{\dag} (\hat \psi_{C})^{-1}},
\end{equation}
as one can see from Eqs. (\ref{C1}) and (\ref{C2}) by virtue of Eqs. (\ref{C6}) and (\ref{C7}).

By exploiting the commutation relation (\ref{C8}), we can write

\begin{equation}
\label{C16}
[\hat \psi_{C}, (\hat \psi_{C}^{\dag})^{-1}]=-(A-B)(\hat \psi_{C}^{\dag})^{-2},
\end{equation}
which can be employed to derive the notable formulae

\begin{equation}
\label{C17}
\hat S=\frac{1}{\sqrt{2}}\sqrt{[2\hat \psi_{C}+(A-B)(\hat \psi_{C}^{\dag})^{-1}](\hat \psi_{C}^{\dag})^{-1}},
\end{equation}
and

\begin{equation}
\label{C18}
\hat S^{\dag}=\frac{1}{\sqrt{2}}\sqrt{\hat \psi_{C}^{-1}[2\hat \psi_{C}^{\dag}+(A-B)\hat \psi_{C}^{-1}]}.
\end{equation}

Equations (\ref{C17}) and (\ref{C18}) are very suitable for a comparative investigation of the SW and Caves' detection.
The operators (\ref{C17}) and (\ref{C18}) reproduce $\hat R$ and $\hat R^{\dag}$ for $A=B$, in correspondence 
of which $\hat \psi_{C} \equiv  \hat \psi_{SW}$ and $\hat \psi_{SW}$, $\hat \psi_{SW}^{\dag}$ commute.

In this case a (feasible) phase makes sense and takes the form 
$\hat \theta_{SW}\equiv \hat \theta=\frac{1}{i} \ln \hat R = \frac{1}{2i}[\ln(\tilde a + \tilde b^{\dag}) 
- \ln (\tilde a^{\dag} + \tilde b)]$.

Moreover, the cosine and sine operators (\ref{Z5}) and (\ref{Z6}) can be meaningfully defined.
The expressions (\ref{C17}) and (\ref{C18}) of the operators $\hat S $ and $\hat S^{\dag}$ represent 
a significant extension of $\hat R $ and $\hat R^{\dag} $ to the noncommutative Caves' heterodyning 
scheme characterized by $[\hat \psi_{C}, \hat \psi_{C}^{\dag}]=(A-B)\hat 1$.

In the noncommutative (Caves) heterodyning, let us introduce the operators

\begin{equation}
\label{C19}
 C_{0}=\frac{1}{2}(\hat S + \hat S^{\dag}),
\end{equation}

\begin{equation}
\label{C20}
S_{0} =\frac{1}{2i}(\hat S - \hat S^{\dag}),
\end{equation}
which possess the properties

\begin{equation}
\label{C21}
[C_{0}, S_{0}]=0,
\end{equation}

\begin{equation}
\label{C22}
C_{0}^{2}+S_{0}^{2}=\hat 1,
\end{equation}
for $A\equiv B$ (SW detection), and

\begin{equation}
\label{C23}
[C_{0}, S_{0}]=\frac{i}{2}
\left[ \sqrt{1+\frac{(A-B)^{2}}{4}(\hat \psi_{C}^{\dag})^{-2}\hat \psi_{C}^{-2}}- 
\sqrt{1+\frac{(A-B)^{2}}{4}\hat \psi_{C}^{-2}(\hat \psi_{C}^{\dag})^{-2}} \right],
\end{equation}

\begin{equation}
\label{C24}
C_{0}^{2}+S_{0}^{2}=\frac{1}{2}
\left[ \sqrt{1+\frac{(A-B)^{2}}{4}(\hat \psi_{C}^{\dag})^{-2}\hat \psi_{C}^{-2}}+ 
\sqrt{1+\frac{(A-B)^{2}}{4}\hat \psi_{C}^{-2}(\hat \psi_{C}^{\dag})^{-2}} \right],
\end{equation}
for $A\neq B$ (Caves' detection).

\subsection{The commutator $[\hat S, \hat N]$}

A remarkable commutation rule can be found for ``noncommutative'' framework extending in the RNS context the 
property $[\hat R, \hat N]=\hat R$ valid for the ``commutative'' case (see Ref. (\ref{N4})).
The generalized  relation can be determined by a direct application of formulae (\ref{N5}) 
and (\ref{H6}) (Ref. \cite{LOUISELL}). This reads

\begin{equation}
\label{C25}
[\hat S, \hat N]=\frac{1-\mu^{2}}{4}\hat S^{-3}\frac{1}{\hat Z^{4}}+
\frac{1}{\hat Z}(\hat S^{-1}\frac{1}{\hat Z}\hat T + \hat T \hat S^{-1}\frac{1}{\hat Z}).
\end{equation}

For $\mu = 1$, Equation (\ref{C25}) reproduces just the RNS representation $[\hat R, \hat N]=\hat R$ which 
holds for the SW framework.

A comment is in order. Our generalization of the operator $\hat R$, leads to an expression for $\hat S$ which is 
not unitary. In fact, it depends on the real parameter $\mu$, and becomes exactly unitary only when $\mu=1$. 
In this limit, $\hat S \equiv \hat R$. Therefore, our extension is consistent with the Shapiro-Wagner theory.
We notice that even if our extended formulae involving $\hat S$ are provided by  {\it exact} relationships, they are 
especially indicated to build up an {\it approximate} RNS theory of quantum measurements of noncommuting observables 
\cite{Y1}-\cite{YL}. This problem, in Caves' context, could give rise to a challenging possible next application.

%%%%%%%%%%%%%%%%%%%%%%%%%%%%%%%%%%%%%%%%%%%%%%%%%%%%%%%%%%%%%%%

\section{Concluding remarks}

The main results achieved in this paper are widely summarized and discussed in the Introduction. 
Therefore, we conclude with a few comments pertinent to some aspects relative to the phase problem.
Since 1927 up to now, several proposals and methods on both quantum phase and amplitude have been put 
forward. Of course, it is difficult to make a satisfactory account of contributions even if these should 
be restricted to a more updated situation. Among many contributions which should be worth mentioning, we 
limit ourselves to quote a sample of them directly connected with the concepts considered in this paper: 
Refs. \cite {KASTRUP}, \cite{RASETTI} and \cite{BGL}. Kastrup \cite{KASTRUP} settles a theoretical background 
yielding a group procedure of quantizing moduli and phases of complex numbers. In Ref. \cite{RASETTI} a fully 
consistent realization of the quantum operators corresponding to the canonically conjugate and number variables 
is carried out. This approach is based on the use of the noncompact Lie algebra $su(1,1)$. Precisely, as a 
mathematical tool Rasetti exploits the $\kappa=\frac{1}{2}$ positive discrete series of the irreducible 
unitary representation of $su(1,1)$ and the double covering group $SO^{\dag}(1,2)$.
The contributions of Kastrup and Rasetti represent interesting attempts to clarify some important aspects 
of the quantum phase problem.

On the other hand, in Ref. \cite{BGL} a systematic reconsideration of the phase problem based on phase 
observables as shift-covariant positive-operator-valued measures is performed. This article deserves to 
be consulted since it yields a coherent unification of several conceptually different approaches to the 
quantum phase problem.

The results of the present paper, whose spirit is essentially speculative, offer some starting points of 
physical applications: one of them concerns a possible detailed investigation on the strategy outlined 
in Section 8 applied to the approximate quantum measurements of generalized observables in Caves' framework.
This program should be pursued in the near future.

\section{Acknowledgements}

It is a pleasure to thank Prof. M. Rasetti of Politecnico di Torino, Italy, for the reading of the manuscript 
and for useful critical comments.


\bigskip

\begin{thebibliography}{99}

\bibitem{NOVEL} A. Geralico, G. Landolfi, G. Ruggeri and G. Soliani, Phys. Rev. D 69, 043504 (2004).
\bibitem{D} P.A.M. Dirac,  Proc. Roy. Soc. (London) A 114, 243 (1927).
\bibitem{CN} P. Carruthers and M.M. Nieto, Rev. Mod. Phy. 40, 411 (1968).
\bibitem{L} R. Lynch, Phys. Rep. 256, 367 (1995).

\bibitem{B1} M. Ban,  Phys. Rev. A 48, 3452 (1993).
\bibitem{B2} M. Ban,  J. Math. Phys. 32, 3077 (1991).
\bibitem{B3} M. Ban, Optics Communications 94, 231 (1992).
\bibitem{HRADIL} Z. Hradil, Phys. Rev. A 47, 2376 (1993).
\bibitem{B4} M. Ban,  Phys. Rev. A 50, 2785 (1994).
\bibitem{CAVES} C.M. Caves, Phys. Rev. D 26, 1817 (1982).
\bibitem{YUEN} H.P. Yuen, Phys. Rev. A 13, 2226 (1976).
\bibitem{LINDNAGEL} G. Lindblad and B. Nagel,  Ann. Inst. Henry Poincare', Vol. XIII, n. 1. 1970, pp. 27-56.
\bibitem{LOUISELL} W.H. Louisell, ``Quantum Statistical Properties of Radiation'', J. Wiley and Sons, 1973.
\bibitem{SW} J.H. Shapiro and S.S. Wagner, IEEE J. Quantum Electron. QE-20, 803 (1984).
\bibitem{RASETTI} M. Rasetti, ``A fully consistent Lie algebraic representation of quantum phase and number operators'', Preprint Politecnico di Torino (2002).
\bibitem{KASTRUP} H.A. Kastrup, How to quantize phase and moduli!, e-print quant-ph/0109013.
\bibitem{BGL} P. Bush, M. Grabowski and P.J. Lathi, Ann. Phys. 237, 1 (1995).

\bibitem{PRATT} W.K. Pratt, ``Laser Communication Systems'', J. Wiley \& Sons, New York, 1969.

\bibitem{EG} C.J. Eliezer and A. Gray, SIAM J. Appl. Math. 30, 463 (1976).
\bibitem{E} V.P. Ermakov, Univ. Izv. Kiev 20, No. 9, 1 (1880).
\bibitem{Mi} E.W. Milne, Phys. Rev. 35, 863 (1930).
\bibitem{Pi} E. Pinney, Proc. Am. Math. Soc. 1, 681 (1950).

\bibitem{Y1} H,P. Yuen, Phys. Lett. 91A, 101 (1982).
\bibitem{YL} H.P. Yuen and M. Lax, IEEE Trans. Inform. Theory 19, 740 (1973).




\end{thebibliography}
\end{document}